\documentclass[preprint,showpacs,preprintnumbers,amsmath,amssymb,pre]{revtex4}

\usepackage{graphicx}
\usepackage{dcolumn}
\usepackage{bm}
\usepackage{color}

\def \Vec#1{\mbox{\boldmath $#1$}}
\def \p{\partial}
\def \f#1#2{\frac{#1}{#2}}
\def \mr#1{\mathrm{#1}}
\def \l{\left}
\def \ri{\right}
\def \ds{\displaystyle}

\begin{document}


\title{Comparison of Entropy Production Rates in Two Different Types of Self-organized Flows: B\'{e}nard Convection and Zonal flow}

\author{Y. Kawazura and Z. Yoshida}
\affiliation{Graduate School of Frontier Sciences, The University of Tokyo,
Kashiwa, Chiba 277-8561, Japan}


\date{\today}

\begin{abstract}
Two different types of self-organizing and sustaining ordered motion in fluids or plasmas ---one is a B\'{e}nard convection (or streamer) and the other is a zonal flow --- have been compared by introducing a thermodynamic phenomenological model and evaluating the corresponding entropy production rates (EP).
These two systems have different topologies in their equivalent circuits: the B\'{e}nard convection is modeled by parallel connection of linear and nonlinear conductances, while the zonal flow is modeled by series connection.
The ``power supply'' that drives the systems is also a determinant of operating modes.
When the energy flux is a control parameter (as in usual plasma experiments), the driver is modeled by a constant-current power supply, and
when the temperature difference between two separate boundaries is controlled (as in usual computational studies), the driver is modeled by a constant-voltage power supply.
The parallel (series)-connection system tends to minimize (maximize) the total EP when a constant-current power supply drives the system.
This minimum/maximum relation flips when a constant-voltage power supply is connected.
\end{abstract}


\maketitle

\section{Introduction}\label{s:Introduction}
In recent studies on non-equilibrium thermodynamics, the entropy production rate (EP) has been highlighted as a promising determinant of self-organized ordered structures.
The minimum EP (MinEP) principle succeeded to describe various structures self-organized in nonlinear dissipative systems~\cite{Prigogine}.
The MinEP principle has its historical origin in Helmholtz' minimum dissipation principle that determines the velocity distribution of a stationary viscous flow~\cite{Helmholtz}. 
L. Onsager generalized this principle to a variational principle relating thermodynamic forces (potentials) and fluxes~\cite{Onsager, Onsager2}. 
The MinEP principle proposed by I. Prigogine, independently rooted in the study of linear discontinuous systems, has extended the scope (by hypothetical studies)  
toward various nonlinear systems.
I. Gyarmati showed the equivalence of Onsager's variational principle and the MinEP principle
in a linear dissipative system~\cite{Gyarmati}.
A literal application of EP in the variational principle, however, may lead to an improper result.
For example, consider a heat diffusion on a homogeneous medium
where Fick's law $\bm{F}=-D\nabla T$ ($\bm{F}$: heat flux, $T$: temperature, $D$: diffusion coefficient) applies.  
Then, the total EP is the volume integral $\int\bm{F}\cdot\nabla T^{-1} dV=\int D|\nabla\log T|^2 dV$, which, however, is not the
target of minimization leading to Fourier's law (which must be $\int D|\nabla T|^2 dV$).

At the opposite pole to the expanded narrative of MinEP,
examples of ``maximizing'' EP have been found in various non-equilibrium systems,
suggesting the possibility of antithetical ``maximum EP (MaxEP)'' principle.
G. W. Paltridge showed, in his pioneering work~\cite{Paltridge}, that the atmospheric heat transfer from the tropical to the polar regions maximizes EP 
in a sense that the temperature contrast tends to be enhanced compared to that of simple heat diffusion; see also~\cite{MEP_ex1,MEP_ex2,MEP_ex3,MEP_ex4}.
MaxEP states are often found in turbulent fluid systems.
Ozawa~\textit{et al.}~\cite{MEP_ex5} pointed out that fluid-mechanical instabilities play an essential role in increasing EP a nonlinear regime.
Yoshida and Mahajan~\cite{Yoshida} formulated a thermodynamic model of heat transfer through a self-organizing fluid (plasma),
demonstrating the bifurcation of a MaxEP operation mode; it is shown that the MaxEP branch is stable 
when the system is ``flux-driven,'' i.e. the heat flux driving the non-equilibrium system is 
the parameter that controls the operation mode;
the opposite is a ``force-driven'' system, in which a thermodynamic 
force (for instance, a temperature difference between two boundaries connected to heat baths) is
the control parameter. 
We note that, in usual plasma experiments (as well as in most natural processes, like atmospheric heat transfer driven by solar heat source), the input power is controlled (or, given as a determining parameter), which must be transferred to some heat sink in a steady state, thus the ``flux'' is the controlling parameter.  On the other hand, theoretical analysis or
a computational study often uses a force-driven model giving a boundary condition on the
intensive quantities (like the temperature).  
Mathematically, a flux-driven condition and a force-driven condition
correspond to a Neumann boundary condition and a Dirichlet boundary condition, respectively
(one may consider a mixed-type boundary condition if appropriate).

Statistical mechanical theories build a foundation for the use of EP as a scalar in variational principle:
R. C. Dewar~\cite{Dewar1, Dewar2} related EP to the count of ``paths'' in phase space by invoking Jaynes' formalism~\cite{Jaynes1, Jaynes2}.
R. K. Niven~\cite{Niven1} proposed a different information-theoretic evaluation of EP;
he also pointed out that the driving condition (what he calls a ``flow controlled'' system is equivalent to the aforementioned flux-driven system) is essential to yield a MaxEP state.

Here, we note that a ``maximum'' EP does not mean the maximization of the dissipation function in a transient/local process; in fact, transient/local EP is bounded only from below.
We say maximum or minimum by comparing the total EP of \emph{bifurcated} operation points in (quasi) stationary states.
In a general nonlinear system, total EP may be changed by a structure (flow in a fluid/plasma system) on a large scale, even if the ultimate dissipation occurs on a small scale (such as the Kolmogorov scale).
The MaxEP state is the operating mode with the largest EP among possible modes of (quasi) stationary states (other authors say ``global maximum/minimum of EP''~\cite{Kleidon, Niven2}).

In view of the rich phenomenologies and various theoretical interpretations, it may be useful to develop a simple paradigm by which one can compare basic structures as well as the modes of operation (or drive) of systems. 
For this purpose, the problem of heat transfer in a plasma provides appropriate examples; a plasma is a strongly nonlinear system which exhibits various aspects of non-equilibrium phenomena. 
As pointed out by Yoshida and Mahajan~\cite{Yoshida}, a typical example of bifurcation of a MaxEP state is the H-mode~\cite{H-mode}, which is stably sustained by a given large-enough heat flux.
Recently formulated variational principle of non-equilibrium thermodynamics~\cite{Kawazura} can determine the stable operating point of a nonlinear system (modeled by a nonlinear impedance), and
evaluate either maximum or minimum EP depending on how the system is driven (or sustained);
the min/max duality is described by a Legendre transformation
of a \emph{generalized dissipation function}.

In the present study, we discuss not only how the system is driven but also how the nonlinear impedance is ``arranged.''
In the H-mode, the nonlinearity of the thermal conduction is caused by the self-organization of a \emph{zonal flow} that 
is directed perpendicular to the heat flow; its strong shear suppresses the turbulence, resulting an increased impedance of heat transport.
However, there is another type of self-organization of an ordered flow
---in fluid mechanics, known as B\'{e}nard convection, and in plasma physics,
called ``streamer''~\cite{streamer}---
which brings about an ``opposite'' effect on heat transport;
such a flow parallels to the direction of heat flow and causes convective heat transport,
resulting in a reduced impedance of heat transport. 
While the zonal flow blocks the thermal conduction, the streamer opens up a new channel of heat transfer.
To model these two different structures by \emph{equivalent circuits},
we consider two different topologies of connecting a nonlinear impedance to a ``baseline'' impedance; 
one is series connection and the other is parallel connection
(here, the baseline is determined by the impedance observed when the macroscopic flow is absent, i.e. \emph{before} the self-organization).
The blocking (series) impedance changes from zero to a finite value as the zonal flow grows and suppresses the
turbulent heat transport (the baseline impedance is that of the turbulent heat transport).
The bypath (parallel) impedance, on the other hand, changes from infinity to a finite value
as the streamer glows (the baseline impedance is that of classical heat diffusion).

The aim of this study is to describe an abstract (general) balance law
that dictates bifurcation, stability, and EP of operation points (quasi-stationary macroscopic states) in a nonlinear driven system; such a thermodynamic relation will be ``mechanism free,''
hence specifying
the mechanisms that yield nonlinear impedances or control which type of flow
self-organizes is not the subject of the present practice.
Here we refer to a limited number of theoretical arguments on relevant 
mechanisms of self-organization.  
The creation of a zonal flow is explained by the so-called
$\beta$ term in the Rossby wave equation\,\cite{Rhines}
and its cousin Hasgawa-Mima equation of drift wave (see a review paper\,\cite{Hasegawa}).
On the other hand, a B\'enard convection is caused by a buoyancy term
measuring a baroclinic effect (which is
related to an inhomogeneity of entropy caused by the gravity; see\,\cite{Landau}).
A ``bifurcation'' of the zonal flow and the streamer has been studied for magnetic-curvature driven Rayleigh-Taylor instabilities, showing that the strength of dissipation is the control parameter\,\cite{Sen}.

\section{Heat engine model of self-organization}
We consider a layer bounded on one side by a high temperature core plasma (temperature $T$) and on the other side by a low temperature outer region (heat bath with temperature $T_0$).
The temperature difference (thermodynamic force) and the heat flux are related by a thermal impedance $Z$ as $T - T_0 = Z F$.
When the system is flux-driven, a heat flux $F$ is given, and then, the inner boundary temperature $T$ is a dependent variable.
On the other hand, in a temperature-driven system, 
the temperature difference $T-T_0$ is given, and $F$ is a dependent variable.

\begin{figure}[tpb]
	\begin{center}
		\includegraphics*[width=0.5\textwidth]{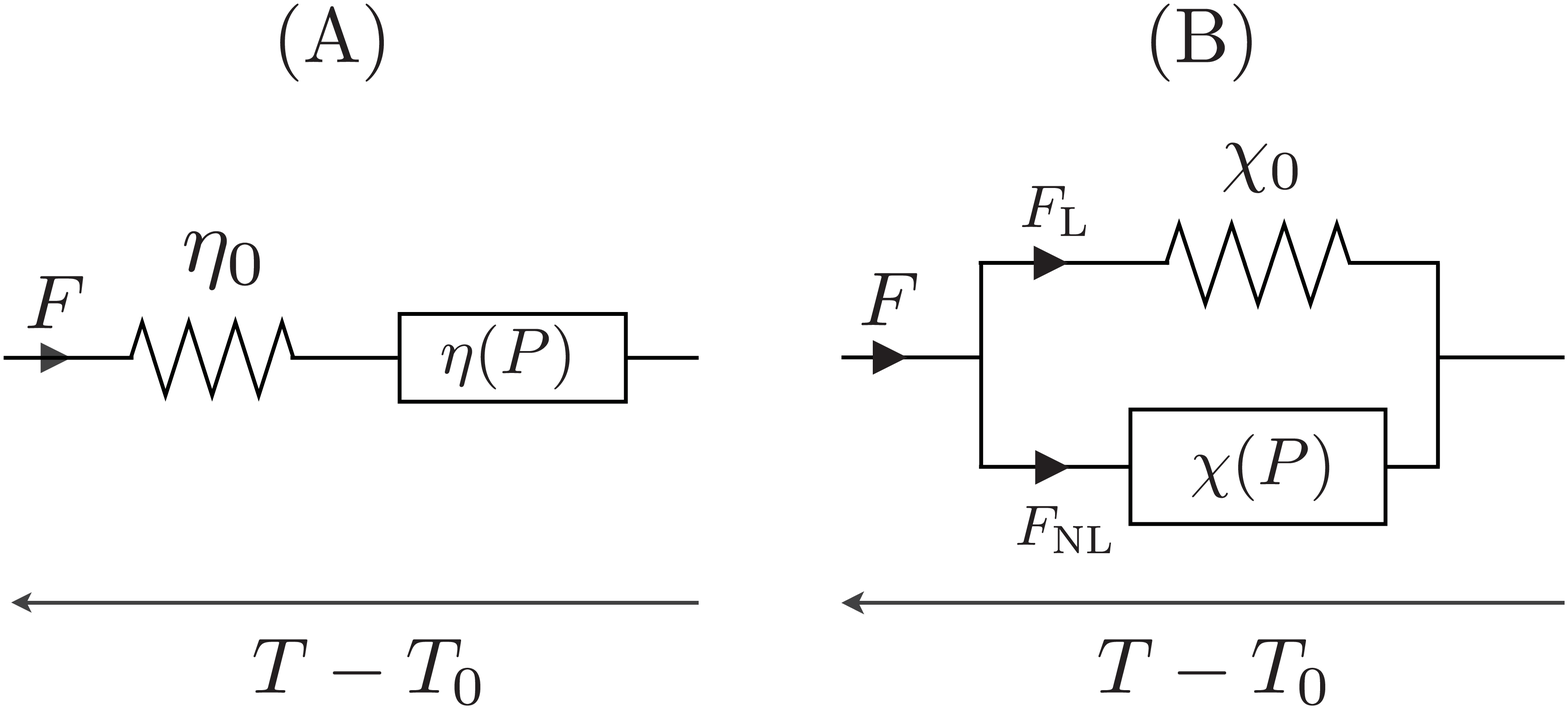}
	\end{center}
	\caption{The equivalent circuits of (A) a zonal-flow and (B) a streamer (B\'{e}nard-convection) models.}
	\label{f:circuit}
\end{figure}

We start by reviewing the thermodynamic model of H-mode/zonal flow self-organization~\cite{Yoshida}.
In a turbulent heat transport, the thermal impedance $Z$ may be written as a sum of two components
$Z = \eta_0 + \eta$; 
$\eta_0$ represents the background (baseline) turbulent heat transport, and $\eta$ 
represents a blocking impedance modeling the suppression of turbulent heat transport by
a zonal flow.
The nonlinear impedance $\eta$ is a function of the available power $P$ to create the flow.
For simplicity, we assume a simple form
\begin{equation}\label{e:eta_P}
	\eta(P) = a P,
\end{equation}
where $a$ is a positive constant modeling the increase of the blocking impedance by creation of a zonal flow. 
The essential nonlinearity of this system is brought about by the power $P$
that obeys the thermodynamic second law.
In an ideal (quasi-stationary) process (Carnot's cycle), the maximum available power as $F(1-T_0/T)$. 
However, the entropy production in the (turbulent) diffusion process diminishes the effective power:
denoting $T_D \equiv T_0 + \eta_0 F$, we estimate
\begin{equation}
	P=F\left(1-\f{T_0}{T}\right)-F\left(1-\f{T_0}{T_D}\right) = F\l( \f{T_0}{T_D} - \f{T_0}{T} \ri).
	\label{e:P_series}
\end{equation}
The force (temperature difference) and flux are related by the impedance:
\begin{equation}\label{e:series operating point}
	T - T_0 = [\eta_0 + \eta(P)]F.
\end{equation}
The operating point is determined by Eqs.~(\ref{e:eta_P}), (\ref{e:P_series}) and (\ref{e:series operating point}).
The equivalent circuit of Eq.~(\ref{e:series operating point}) is shown in Fig.~\ref{f:circuit} (A).

Next we formulate a model of streamer (B\'{e}nard convection).
Since a streamer creates a new route of heat conduction in parallel to the ambient (baseline)
diffusion path, we consider a parallel-connection circuit such as Fig.~\ref{f:circuit} (B).
In terms of a baseline conductance $\chi_0$ (assumed to be a constant) and a nonlinear conductance $\chi(P)$, 
the flux $F$ and the temperature $T$ are related by
\begin{equation}\label{e:parallel operating point}
	F = [\chi_0 + \chi(P)](T-T_0).
\end{equation}
The heat flow is (conceptually) divided into two channels; $F_\mathrm{L}=\chi_0(T-T_0)$ in the baseline conductance and $F_\mathrm{NL}=\chi(P)(T-T_0)$ in the nonlinear conductance; see Fig.~\ref{f:circuit} (B).
%

We assume that the nonlinear conductance $\chi(P)$ is written as 
\begin{equation}\label{e:chi_P}
	\chi(P) = b P + c P^2,
\end{equation}
where $b$ and $c$ are constants.
The reason why we need the second-order term $cP^2$ will be explained later.
The power $P$ is determined by (\ref{e:P_series}); here we rewrite it as
\begin{equation}
	P = F\l( 1 - \f{T_0}{T}\ri) - F_\mathrm{L}\l( 1 - \f{T_0}{T}\ri) = \l( 1 - \f{T_0}{T} \ri)[ F - \chi_0(T-T_0) ].
	\label{e:P_parallel}
\end{equation}
The operating point is determined by Eqs.~(\ref{e:parallel operating point}), (\ref{e:chi_P}) and (\ref{e:P_parallel}).

\begin{figure}[tpb]
	\begin{center}
		\includegraphics*[width=0.45\textwidth]{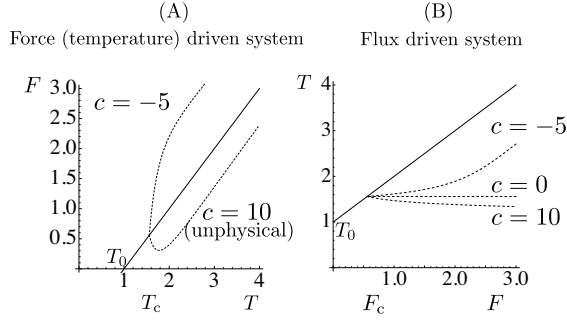}
	\end{center}
	\caption{The operating points of a nonlinear parallel circuit connected to (A) a constant-force (temperature) drive, and (B) a constant-flux drive. 
	The linear solutions are displayed by solid lines, and the nonlinear solutions are, by dashed lines. Here we assume $T_0 = 1, \chi_0 = 1, b = 5$ and $c = (-5, 0, 10)$.
	In (A) the nonlinear branch with $c=10$ is unphysical, since the corresponding power $P$ becomes negative.}
	\label{f:operating point}
\end{figure}

One may solve the series model Eq. (\ref{e:series operating point}) or the parallel model Eq. (\ref{e:parallel operating point}) either for $T$ (with respect to a given $F$) or $F$ (with respect to a given $T$).
Since the series model has been analyzed in \cite{Yoshida}, we focus on the parallel model.
Solving Eqs.~(\ref{e:parallel operating point}), (\ref{e:chi_P}) and (\ref{e:P_parallel}) for $F$, we obtain two branches:
\begin{eqnarray}\label{e:sol_F}
	F=
	\l\{
		\begin{array}{lll}
			F_1 \equiv & \ds \l(T-T_0\ri) \chi _0 & (T\le T_\mathrm{c})\\
			~\\
			F_2 \equiv & \ds \f{1}{c\l(T - T_0\ri)^3}\l[c\chi_0T^4 - \l(b + 4cT_0\chi_0\ri)T^3 \ri.\\
										 &\l. + \l(1 + 2bT_0 + 6cT_0^2\chi_0\ri)T^2 - \l(bT_0^2 + 4cT_0^3\chi_0\ri)T + cT_0^4\chi_0\ri] & (T\ge T_\mathrm{c})\\
		\end{array}
	\ri.,
\end{eqnarray}
where 
\begin{equation}
	T_\mathrm{c} = \f{1 + 2bT_0 + \sqrt{1 + 4bT_0}}{2b}
	\label{e:T_c}
\end{equation}
is threshold temperature at which the bifurcation occurs.
Evidently, $F = F_1$ corresponds to baseline \emph{linear state}, while $F = F_2$ (achieved by $P>0$) represents the \textit{nonlinear state} with a streamer; see Fig.~\ref{f:operating point}~(A).

By these solutions, we find that the critical temperature $T_\mathrm{c}$ is determined only by $b$.
The bifurcation of the nonlinear solution occurs for every positive $b$; this is in marked contrast to the zonal-flow (series) model in which
the bifurcation needs a sufficiently large $a$~\cite{Yoshida}.
We also find that the nonlinear solution becomes singular if $c=0$;
a higher order nonlinearity of $\chi(P)$ determines $F_2$.
In the temperature-driven system, the nonlinear solution in the regime of $c>0$ is unphysical, because the condition $F_2<\chi_0(T - T_0)$ makes the power $P$ negative; see Eq.~(\ref{e:P_parallel}).

\begin{figure}[tpb]
	\begin{center}
		\includegraphics*[width=0.43\textwidth]{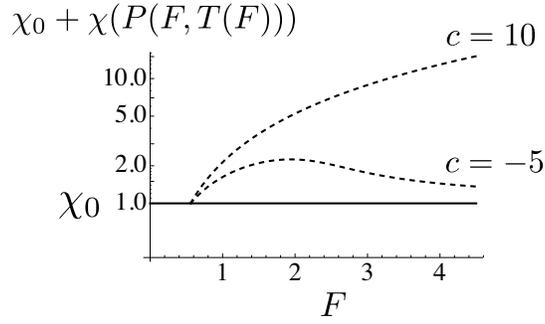}
	\end{center}
	\caption{The conductance $\chi_0 + \chi(P(F, T(F)))$ of the flux-driven system, plotted as a function of the heat flux $F$.}
	\label{f:chi}
\end{figure}

The flux-driven system is more complicated because the nonlinear solution $T = T_2(F)$ is given as an implicit function.
In Fig.~\ref{f:operating point}, we plot the numerical solution of the operating points.
In the temperature-driven system the nonlinear solution must enhance the heat flux, thus we have to assume $c < 0$. 
While the force-driven system has physical solution only for $c < 0$, the flux-driven system has physical solutions for every $c$, satisfying $T < T_0 + F/\chi_0$.
If $c<0$, $T$ increases monotonically with $F$. The temperature difference diminishes when $c$ approaches 0, and, $T = T_\mathrm{c}$ (constant) for all $F$ when $c = 0$.
With a positive $c$, $T$ decreases monotonically with the increase in $F$, and approaches $T_0$.
In Fig.~\ref{f:chi}, we plot the conductance $\chi_0 + \chi(P(F,T(F)))$ of the flux-driven system as a function of $F$.
If $c<0$, the nonlinear conductance converges to $\chi_0$ in the limit of $F \to \infty$.

\section{Stability}
We analyze the stability of each branch, and determine which branch is realized.
We start by the temperature-driven system.
We fix the temperature $T$ and define the total conductance $C \equiv \chi_0 + \chi(P)$. 
The following chain of events is caused by a fluctuation $\delta F$ in the heat flux: 
\begin{displaymath}
  \begin{array}{lll}
        \delta F & \rightarrow &  \delta C =\displaystyle  \f{\p  C}{\p F}\delta F \\
                \\
                & \rightarrow &  \delta F' = (T-T_0)\displaystyle  \f{\p C}{\p F}\delta F \equiv \alpha^{T}\delta F.  
	\end{array}
\end{displaymath}
The factor $\alpha^T$ scales the amplification of fluctuation $\delta F$. 
The system is stable (unstable) if $\alpha^T < 1$ ($\alpha^T > 1$).
The first step of the events requires certain amount of time while other events can occur
simultaneously.
When $\Delta t$ is the time period of one cycle, the flux perturbation grows or diminishes as $\delta F_n := \delta F|_{t=n\Delta t} = \alpha^n \delta F_0$.
The amplification factor of linear solution is estimated as 
\begin{equation}
	\alpha^T_1 = b(T - T_0)\l( 1 - \f{T_0}{T} \ri).
\end{equation}
On the other hand, the amplification factor of the nonlinear solution is 
\begin{equation}
	\alpha^T_2 = 2 - \f{b(T - T_0)^2}{T}.
\end{equation}
Evidently, $\alpha^T_1<1$ for $T<T_{\mr{c}}$, while $\alpha^T_1>1$ and $\alpha^T_2<1$ for $T>T_{\mr{c}}$. 
Therefore, the linear solution, which is the unique and stable solution in $T<T_{\mr{c}}$, is destabilized at the bifurcation point $T_{\mr{c}}$, and, beyond the bifurcation, the nonlinear branch takes over the linear (smaller $F$) solution as the stable branch. 

\begin{figure}[tpb]
	\begin{center}
		\includegraphics*[width=0.43\textwidth]{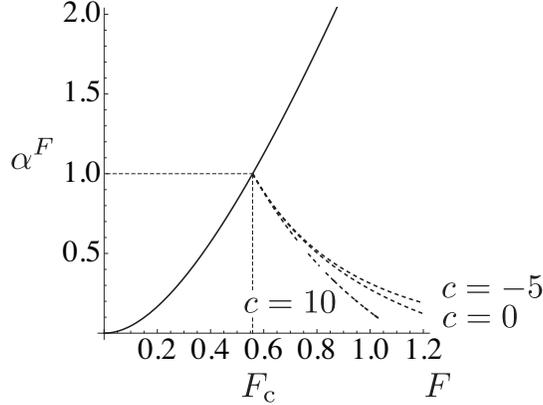}
	\end{center}
	\caption{Thermodynamic stability of the flux-driven parallel (streamer) system. The linear branch (solid line) is stable in the regime of $F<F_{\mr{c}}$ (the critical flux of bifurcation). 
	In $F>F_{\mr{c}}$, the nonlinear branch (dashed lines) with all $c$ is stable, while the linear branch is unstable.}
	\label{f:stability}
\end{figure}

Next, we study the stability of the flux-driven system. 
In this case, it is difficult to display the amplification factor explicitly.
Here we consider the chain of events caused by a fluctuation $\delta T$ in the temperature: 
\begin{displaymath}
  \begin{array}{lll}
        \delta T & \rightarrow &  \delta Z =\displaystyle  \f{\p  Z}{\p T}\delta T \\
                \\
                & \rightarrow &  \delta T' = T\displaystyle  \f{\p Z}{\p T}\delta T \equiv \alpha^{F}\delta T.  
				\end{array}
\end{displaymath}
We plot the numerical estimate of the amplification factor of each branch in Fig.~\ref{f:stability}.
The linear solution destabilizes beyond the critical flux $F_{\mr{c}}$, and the bifurcated nonlinear branch is stable for every value of $c$.

\section{Comparison of EP in parallel and series systems}
\begin{figure}[htpb]
	\begin{center}
		\includegraphics*[width=0.5\textwidth]{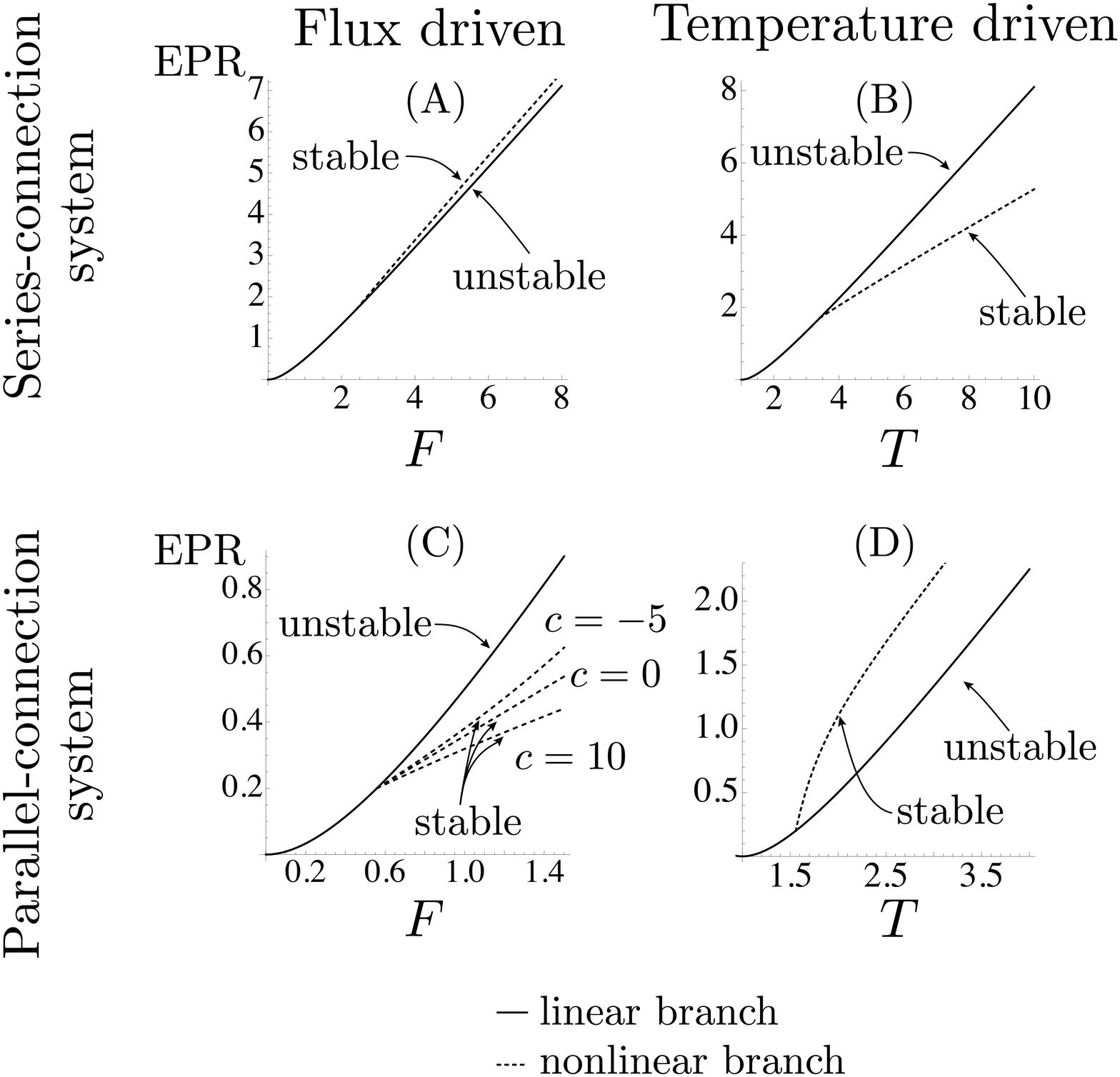}
	\end{center}
	\caption{The entropy production rate (EP) of the series/parallel-connection systems in flux-driven/temperature-driven operations, which is either maximized or minimized, by the nonlinear branch of solution (that is always the stable branch when it exists), depending on the topology of connection (series or parallel), as well as how the system is driven.}
	\label{f:EP}
\end{figure}

Here, we compare EP in parallel/series systems connected to a flux/temperature drives.
The entropy production caused by heat flow is given by the volume integral $\int \Vec{F}\cdot\nabla(1/T)dV$, where $\Vec{F}$ is the local heat flux. 
In a stationary system, $\nabla\cdot \Vec{F} = 0$, thus, integrating by parts, we may represent EP by the boundary terms as $F(1/T_0 - 1/T)$.
Figure~\ref{f:EP} shows EP in the flux-driven and temperature-driven systems.
In the series-connection system, EP is maximized by the nonlinear solution when it is flux-driven (A), but is minimized when it is temperature-driven (B).
On the contrary, EP of the parallel-connection system is minimized  by the nonlinear solution when it is flux-driven (C), and is maximized when it is temperature-driven (D).
R. K. Niven~\cite{Niven2} studied numerically the EP of turbulent flow in parallel pipes driven by a fixed head (force), and found similar behavior.

%

Table~\ref{t:EP} summarizes the EP of the series-connection and the parallel-connection systems in flux-driven and temperature-driven operations. 

\begin{table}[tb]
	\caption{EP of the series-connection and the parallel-connection systems in flux-driven and temperature-driven modes of operation.}
	\begin{tabular}{|c|c|c|}
		\hline
		 & Flux-driven & Temperature-driven\\
		\hline
		Series-connection system & Max & Min\\
		\hline
		Parallel-connection system & Min & Max\\
		\hline
	\end{tabular}
	\label{t:EP}
\end{table}

\section{Conclusion}
We have introduced a paradigm of describing operating points (macroscopic quasi-stationary state)
of a nonlinear heat conduction system; 
two different connections (series and parallel connections) of nonlinear impedances (modeling the self-organization of zonal flow and streamer, respectively), as well as two different types of drives (flux drive and temperature drive),
frame the paradigm.  Thermodynamic stability of bifurcated operating points
can be understood by the phenomenological model based on this paradigm;
the nonlinear branch (created by a positive power $P$) is always stable (when it bifurcates from the linear branch).
The parallel and series-connection systems, respectively, minimize and maximize EP when connected to a flux drive.
When they are connected to a temperature-drive, however, the minimum/maximum relation flips.

\end{document}